\documentclass[a4paper]{spie}  

\usepackage[]{graphicx}
\usepackage{psfrag}

\title{LOPES \\ Detecting Radio Emission from Cosmic Ray Air Showers} 

\author{\noindent 
A.~Horneffer$^{a,*}$,
T.~Antoni$^{b}$, 
W.D.~Apel$^{c}$, 
F.~Badea$^{b,1}$,
K.~Bekk$^{c}$, 
A.~Bercuci$^{c,1}$,
M.~Bertaina$^{d}$,
H.~Bl\"umer$^{c,a}$,
H.~Bozdog$^{c}$,
I.M.~Brancus$^{e}$,
M.~Br\"uggemann$^{f}$,
P.~Buchholz$^{f}$,
C.~B\"uttner$^{b}$,
A.~Chiavassa$^{d}$,
K.~Daumiller$^{b}$, 
C.M.~de Vos$^{g}$,
P.~Doll$^{c}$, 
R.~Engel$^{c}$,
J.~Engler$^{c}$, 
H.~Falcke$^{g}$,
F.~Fe{\ss}ler$^{c}$, 
P.L.~Ghia$^{h}$,
H.J.~Gils$^{c}$,
R.~Glasstetter$^{i}$, 
A.~Haungs$^{c}$, 
D.~Heck$^{c}$, 
J.R.~H\"orandel$^{b}$, 
T.~Huege$^{a}$,
K.-H.~Kampert$^{i}$,
G.W.~Kant$^{g}$,
H.O.~Klages$^{c}$, 
Y.~Kolotaev$^{f}$,
G.~Maier$^{c}$,
H.J.~Mathes$^{c}$, 
H.J.~Mayer$^{c}$, 
J.~Milke$^{c}$, 
C.~Morello$^{h}$,
M.~M\"uller$^{c}$, 
G.~Navarra$^{d}$,
R.~Obenland$^{c}$,
J.~Oehlschl\"ager$^{c}$, 
S.~Ostapchenko$^{b,2}$, 
M.~Petcu$^{e}$, 
S.~Plewnia$^{c}$, 
H.~Rebel$^{c}$, 
A.~Risse$^{j}$, 
M.~Roth$^{b}$, 
H.~Schieler$^{c}$, 
J.~Scholz$^{c}$, 
M.~St\"umpert$^{b}$, 
T.~Thouw$^{c}$, 
G.C.~Trinchero$^{h}$,
H.~Ulrich$^{c}$,
S.~Valchierotti$^{d}$,
J.~van~Buren$^{c}$,
W.~Walkowiak$^{f}$,
A.~Weindl$^{c}$,
J.~Wochele$^{c}$, 
J.~Zabierowski$^{j}$,
S.~Zagromski$^{c}$\\ 
{\em
\noindent
\skiplinehalf 
$^{a}$ Max-Planck-Institut f\"ur Radioastronomie, 53121 Bonn, Germany\\
$^{b}$ Institut f\"ur Experimentelle Kernphysik, Universit\"at Karlsruhe, 76021 Karlsruhe, Germany,\\
$^{c}$ Institut\ f\"ur Kernphysik, Forschungszentrum Karlsruhe, 76021~Karlsruhe, Germany\\
$^{d}$ Dipartimento di Fisica Generale dell'Universit{\`a}, 10125 Torino, Italy\\
$^{e}$ National Institute of Physics and Nuclear Engineering,7690~Bucharest, Romania\\
$^{f}$ Fachbereich Physik, Universit\"at Siegen, 57072 Siegen, Germany \\
$^{g}$ ASTRON, 7990 AA Dwingeloo, The Netherlands\\
$^{h}$ Istituto di Fisica dello Spazio Interplanetario, CNR, 10133 Torino, Italy \\
$^{i}$ Fachbereich Physik, Universit\"at Wuppertal, 42097 Wuppertal, Germany \\
$^{j}$ Soltan Institute for Nuclear Studies, 90950~Lodz, Poland\\[1.5ex]
$^{1}$ on leave of absence from Nat.\ Inst.\ of Phys.\ and 
Nucl.\ Engineering, Bucharest, Romania\\
$^{2}$ on leave of absence from Moscow State University, 
119899~Moscow, Russia\\
}
}

\authorinfo{$^*$Corresponding author: A.~Horneffer {\em horneff@MPIfR-Bonn.mpg.de}}

\begin{document} 
  \maketitle 

\begin{abstract}
Radio pulses emitted in the atmosphere during the air shower development
of high-energy primary cosmic rays were measured during the late
1960ies in the frequency range from 2 MHz to 520 MHz. Mainly due to
difficulties with radio interference these measurements ceased in the late
1970ies.

LOFAR ({\bf Lo}w {\bf F}requency {\bf Ar}ray) is a new digital radio 
interferometer under
development. Using high bandwidth ADCs and fast data processing it will be 
able to filter out most of the interference. By storing the whole waveform
information in digital form one can analyze transient events like air showers 
even after they have been recorded.

To test this new technology and to demonstrate its ability to measure air
showers
a ''LOFAR Prototype Station'' (LOPES) is set up to operate in conjunction
with an existing air shower array (KASCADE-Grande).

The first phase consisting of 10 antennas is already running. It operates in 
the
frequency range of 40 to 80 MHz, using simple short dipole antennas and direct 
2nd Nyquist sampling of the incoming wave.
It has proven to be able to do simple astronomical measurements, like imaging 
of a solar burst.
It has also demonstrated how 
digital interference suppression and beamforming can overcome the problem of 
radio interference and pick out air shower events.

\end{abstract}



\section{Introduction}

The earth's atmosphere is continuously bombarded by high energy particles, 
called cosmic rays.
They consist mainly of ionized atomic nuclei, of elements in the range 
from  hydrogen to iron and that have energies from $10^{3}$~eV up to the order 
of $10^{20}$~eV. The flux spectrum follows a power 
law $dN/dE \propto E^{-\gamma}$ with a change of the index 
from $\gamma \approx 2.7$ to $\gamma \approx 3.1$ around 3~PeV.
The origin of cosmic rays is still not definitively known. Possible sources 
are supernova remanents, pulsars, quasars, active galactic nuclei, or even 
the decay of topological defects or other exotic particles. Due to 
the wide range in energy different emission mechanisms  are also possible.
Since these charged particles are deflected in the interstellar magnetic 
fields, important information about their origin is in their energy and mass.

High energy cosmic rays hitting the earth's atmosphere undergo nuclear 
reactions with atmospheric nuclei, producing secondary particles in an air 
shower.
Thus direct measurements are only possible above the earth (i.e. with 
balloon-borne or satellite experiments). Due to the low flux cosmic rays at 
energies above ca. 1~PeV cannot be effectively measured by direct measurements.
A standard method to observe these cosmic rays is to measure the
secondary particles of an air shower with an array of particle detectors 
on the ground.
As the state of an air shower at the ground level depends on many factors, 
like primary particle energy and type, atmospheric conditions and statistical 
fluctuations, 
the determination of primary particle energy and type from the 
measured particles is not straightforward.
Very useful information for the determination of primary particle energy and 
type can be obtained by additionally observing the air shower as it evolves.
So far this is only done by observing optical emission like Cherenkov or 
fluorescence light. This requires dark, clear and moonless nights and thus
limits the available duty cycle to about 10\%.

Measuring radio emission from air showers might be an alternative method for
such observations, providing a much better efficiency.
This becomes particularly relevant since a new generation of digital radio
telescopes 
-- designed primarily for astronomical purposes -- 
promises a new way of measuring air showers.

\section{Radio Properties of Air Showers}

\begin{figure}
\begin{center}
   \psfrag{nu0MHz}[c][B]{$\nu$~[MHz]}
   \psfrag{Eomega0muVpmpMHz}[c][t]{$\left|\vec{E}(\vec{R},\omega)\right|$~[$
\mu$V~m$^{-1}$~MHz$^{-1}$]}
\psfrag{Enu0muVpmpMHz}[c][b]{}
   \includegraphics[width=0.6\linewidth]{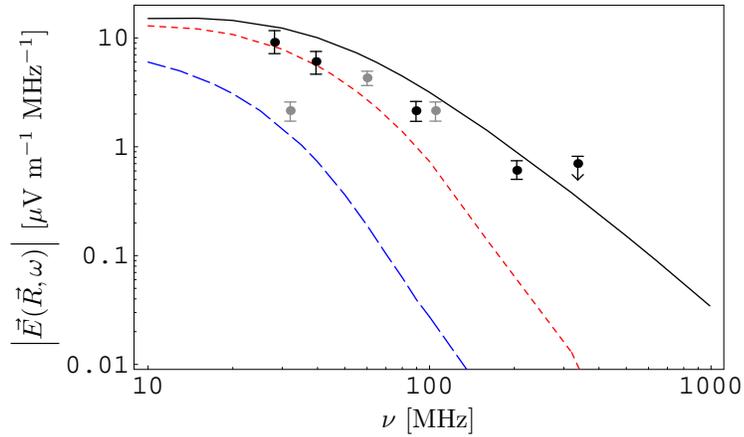}
\end{center}
   \caption{
   \label{fig:spectra_data}
Spectrum of the electric field strength for a vertical $10^{17}$~eV air shower\protect\cite{Huege2003}
Solid: center of illuminated area,
short-dashed: 100~m from center, long-dashed: 250~m from center, 
black points: data as presented by Allan\protect\cite{Allan1971}, 
grey points: data from Prah\protect\cite{Prah1971} 
   }
\end{figure}

Radio emission from cosmic ray air showers were discovered for the first time
by Jelley et al.\cite{Jelley1965}
at 44\,MHz. The results were soon verified and in the 
late 1960's emission from 2\,MHz up to 520\,MHz were found. 
In the following years these activities ceased due to 
difficulty with radio interference, uncertainty about the 
interpretation of the results, and the success of other methods.

The radio properties of extensive air showers were summarized in a
review by Allan\cite{Allan1971}.
His analysis led to 
an approximate formula relating the received voltage to 
various other parameters:
\begin{equation}
 \epsilon_{\nu}   =  20 
\left(\frac{\rm E_p}{10^{17}{\rm eV}}\right ) \sin \alpha \cos \theta   
 \exp\left(\frac{\rm -R}{{\rm R}_0 (\nu,\theta)}\right) \left[\frac{\mu {\rm V}}{\rm m \, MHz}\right]
\end{equation}
Here ${\rm E_p}$ is the primary particle energy, $\alpha$ is the angle to the
geomagnetic field, $\theta$ is the zenith angle, R is the distance to 
the shower center, ${\rm R_0}$ is around 110\,m at 55\,MHz, and $\nu$ is the 
observing frequency. 
Some later works (e.g. by Prah\cite{Prah1971}) yielded much smaller values for
the field strength. Part of this discrepancies may be due to errors in the 
calibration (e.g. of the primary particle energy). Additionally the 
documentation of the available data is not always precise, which makes direct
comparisons complicated. Although comparisons of the absolute values are 
difficult, the trend in the dependence on observing frequency and radial 
distance is fairly consistent.
More recent experiments conducted at the CASA-MIA air shower array have only 
been able to give an upper limit for the field strength\cite{GreenRosnerSuprun2003}.

Recent theoretical studies\cite{FalckeGorham2003}\cite{Huege2003}, modeling 
the radio emission from 
air showers as synchrotron radiation in the earth's magnetic field, have been 
able to reproduce the existing data to a good degree (see 
fig~\ref{fig:spectra_data}).

\section{LOFAR and LOPES}

LOFAR is a new attempt to revitalize astrophysical research at 10-200\,MHz
with the means of modern information technology\cite{Bregman99}.
The basic idea of LOFAR is to build a large array of $100$ stations of 
$100$ omnidirectional dipole antennas in which the received waves are digitized and sent to 
a central super-cluster of computers. 

A new feature is the possibility to store  the entire data stream for a certain
period of time.  If one detects a transient phenomenon 
-- like gamma ray bursts, X-ray binary flares or air showers -- 
one  can then retrospectively form a beam in the desired direction
and thus basically look back in time. 
LOFAR therefore combines the advantages of a 
low-gain antenna (large field of view) and a high-gain antenna (high 
sensitivity and background suppression). This makes it an ideal tool to study
radio emission from cosmic ray air showers. 
With its range of baselines between 10\,m and 400\,km
LOFAR will be capable to detect air showers from $>2\cdot10^{14}$\,eV to 
$\sim10^{20}$\,eV.

To test the technology of LOFAR and demonstrate its capability to measure 
air showers we are building LOPES 
a ''{\bf LO}FAR {\bf P}rototyp{\bf e} {\bf S}tation''
at the site of KASCADE-Grande in Karls\-ruhe/Germany\cite{KASCADE2003}\cite{Schieler02}.
The data from a well tested air shower experiment not only allows us to 
calibrate the radio data with other air shower parameters, it also provides us 
with starting points for the air shower reconstruction, simplifying the 
development process.
This will enable us to clarify the nature and properties of radio emission 
from air showers and provide an energy calibration for future radio air shower 
experiments.
Also, LOPES will provide KASCADE-Grande with valuable additional information 
about the air shower, as the radio data and the particle data come from 
different stages in the evolution of a shower.

\begin{figure}
\begin{center}
\includegraphics[width=0.8\linewidth]{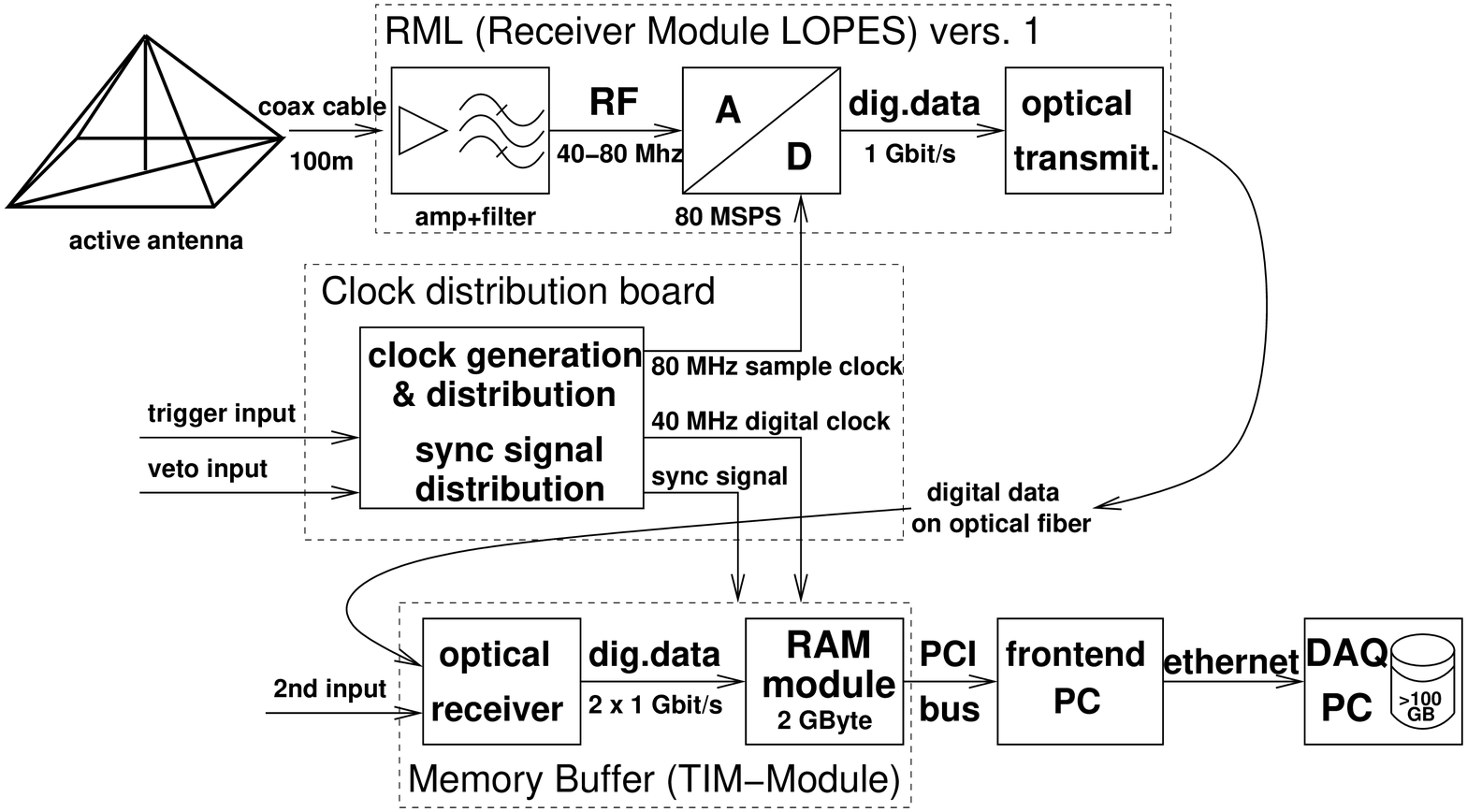}
\caption{
 \label{fig:Hardware-outline}
Outline of the hardware of the first LOPES stage.
   }
\end{center}
\begin{center}
\includegraphics[width=0.45\linewidth]{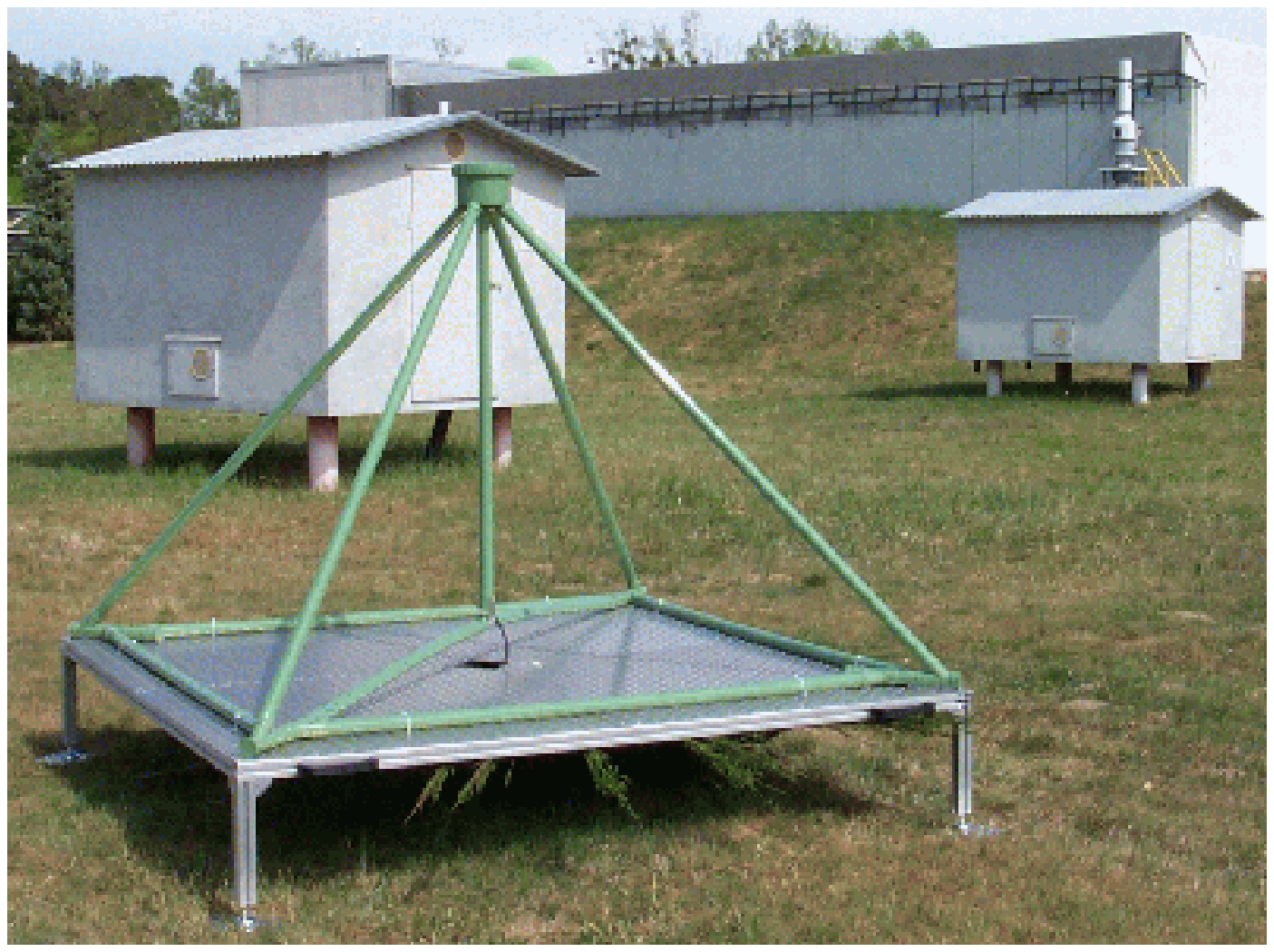}
\caption{
 \label{fig:Antenna}
One of the LOPES antennas at the KASCADE-Grande site. The active balun resides 
inside the container at the top of
the antenna. The radiator consists of cables in two opposing edges of the 
pyramid. By choosing the east-west edges our 
antenna is sensitive to the east-west polarized component of the radiation.
   }
\end{center}
\end{figure}

\section{The Hardware of LOPES}

LOPES operates in the frequency range of 40--80\,MHz. This is a band where 
there are few strong man made radio transmitters, as it lies between the 
short-wave- and the FM-band. Also the frequency is low enough, that the
emission from air showers is strong, while it is still high enough, so that the
background emission from the galactic plane\footnote{The radiation temperature
averaged over the sky ranges from $\sim$2000~K at 80~MHz to $\sim$10000~K at 40~MHz} is 
still low.

The outline of the hardware used for LOPES can be seen in 
figure~\ref{fig:Hardware-outline}. It samples the radio frequency signal after
minimal analog treatment without the use of a local oscillator. This will allow
mass production at low costs in the future.

\subsection{Antenna}

The antennas for LOPES are short dipole antennas with an ''inverted vee'' 
shape. Placed above a suitable ground screen this gives an antenna diagram, 
that points to the zenith and has its half power line at a zenith angle of 
ca. $45^{\circ}$ almost independent of azimuth.
One of the LOPES antennas at the KASCADE-Grande site is shown in
figure~\ref{fig:Antenna}. The visible parts are commercial PVC pipe holding the
active parts in place. The radiator consists of two copper cables extending 
from the top down two thirds of two opposing edges of the pyramid.
The four edges can be used for two orthogonal linear polarizations of the 
signal. As we expect the signal to be highly polarized we use only the 
east-west polarization direction. 

Inside the container at the top resides the active balun. Its main functions 
are balanced to unbalanced conversion, amplification of the signal and 
transformation of the antenna impedance to the $50~\Omega$ impedance of the 
cable. The amplifier is a negative feedback amplifier, with the input 
impedance of the feedback network matched to the impedance of the radiator.
This gives sensitivity over a wide frequency range with good linearity and 
noise performance.

The PVC exterior of the antenna resides on an aluminum pedestal. This acts as 
a ground screen and protects the antenna from the people mowing the lawn.

\subsection{Receiver Module}

\begin{figure}
\begin{center}
\includegraphics[height=0.55\linewidth,angle=-90]{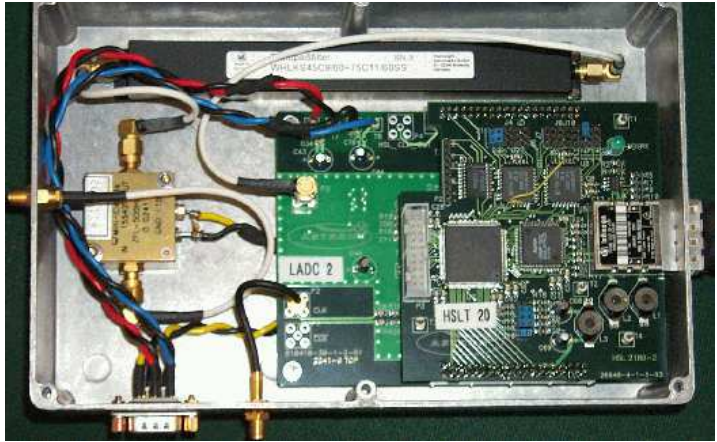}
\caption{
 \label{fig:RML}
A LOPES receiver module with amplifier (left), anti-aliasing filter (top) 
and A/D-converter board with an optical transmitter board piggybacked onto it 
(bottom right).  }
\end{center}
\end{figure}

Figure~\ref{fig:RML} shows one of the receiver modules for LOPES (RML).
On the left side, next to the signal input, is an analog amplifier. This is a
commercially available amplifier. The high background from 
the galactic noise and the low absolute power makes it possible the meet the
required noise and intermodulation performance with a relatively cheap 
amplifier.

Next in the signal path, at the top of picture, is the 
anti-aliasing filter. To suppress contamination from outside our band a 
stopband attenuation of 60~dB is needed. Additionally the desire for high usable 
bandwidth makes steep edges necessary. The filter used for LOPES gives us a 
usable frequency band from 43~MHz to 76~MHz.

The last analog device in the signal path is the A/D-converter board at the
bottom right of the picture. The necessary dynamic range to detect weak pulses 
while not saturating the ADC with radio interference is achieved by using 
12-bit ADCs. We are using ADCs running at 80~MHz, thus sampling the signal
in the 2nd Nyquist domain of the ADCs. Piggybacked onto the A/D-converter 
board is an optical transmitter board for transmission of the digital data to 
the backend module.

\subsection{Digital Backend and Clock Module}

The digital data is transferred via fiber optics to memory modules. These 
modules have standard PCI-connectors ant fit into the front-end PCs. Each 
module has two inputs with $\sim$1.2~GBit/sec each and can take up to 2~GByte 
of normal PC133--style computer memory. This allows it to store up 
to 6.25~seconds of  data from both inputs or even 12.5~seconds of data using 
only one input.
Several of these modules can be used together by synchronizing them with a 
common sync-signal. The modules can either start writing the data after a 
sync-signal or write data continuously into the memory and stop a predefined 
time after a sync-signal is received.

The sample clock for the A/D-converters and a synchronous clock for the memory 
modules is generated on a central clock module and then distributed all 
modules. This clock module also distributes the sync-signal to all memory 
modules.

\section{LOPES at KASCADE-Grande}

\begin{figure}
\begin{center}
\includegraphics[width=0.6\linewidth]{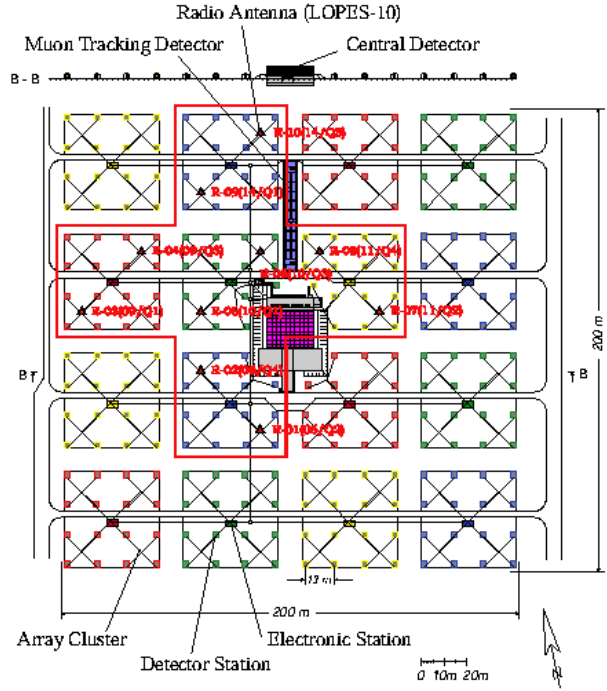}
\caption{
 \label{fig:KASCADE-layout}
Placement of the first 10 LOPES antennas inside the KASCADE-array at KASCADE-Grande.
   }
\end{center}
\end{figure}

The first stage of LOPES is complete and running. The system and ten antennas 
are set up at the  KASCADE-Grande site at the positions shown in 
Figure~\ref{fig:KASCADE-layout}. The relative positions of the antennas have
been measured using a differential GPS system with an accuracy of a few cm.

LOPES is triggered by a large event trigger from the KASCADE-array\cite{Schieler02} (10 out of
16 array-cluster have an internal trigger). 
After each trigger 0.82~milliseconds worth of data from the time around the 
trigger are read out and stored on harddisk in our central DAQ-PC. 

LOPES is not yet tied into the KASCADE data acquisition system. But offline 
correlation of LOPES and KASCADE data works without problems.
A PCI card, that supplies a KASCADE-style timestamp to LOPES, is currently
under development. This timestamp is needed by the event building software of 
KASCADE, and will enable us to add the LOPES data into the KASCADE data stream.

\section{Measurements}

\subsection{Dynamic Spectra}

\begin{figure}
\begin{center}
\includegraphics[width=0.69\linewidth]{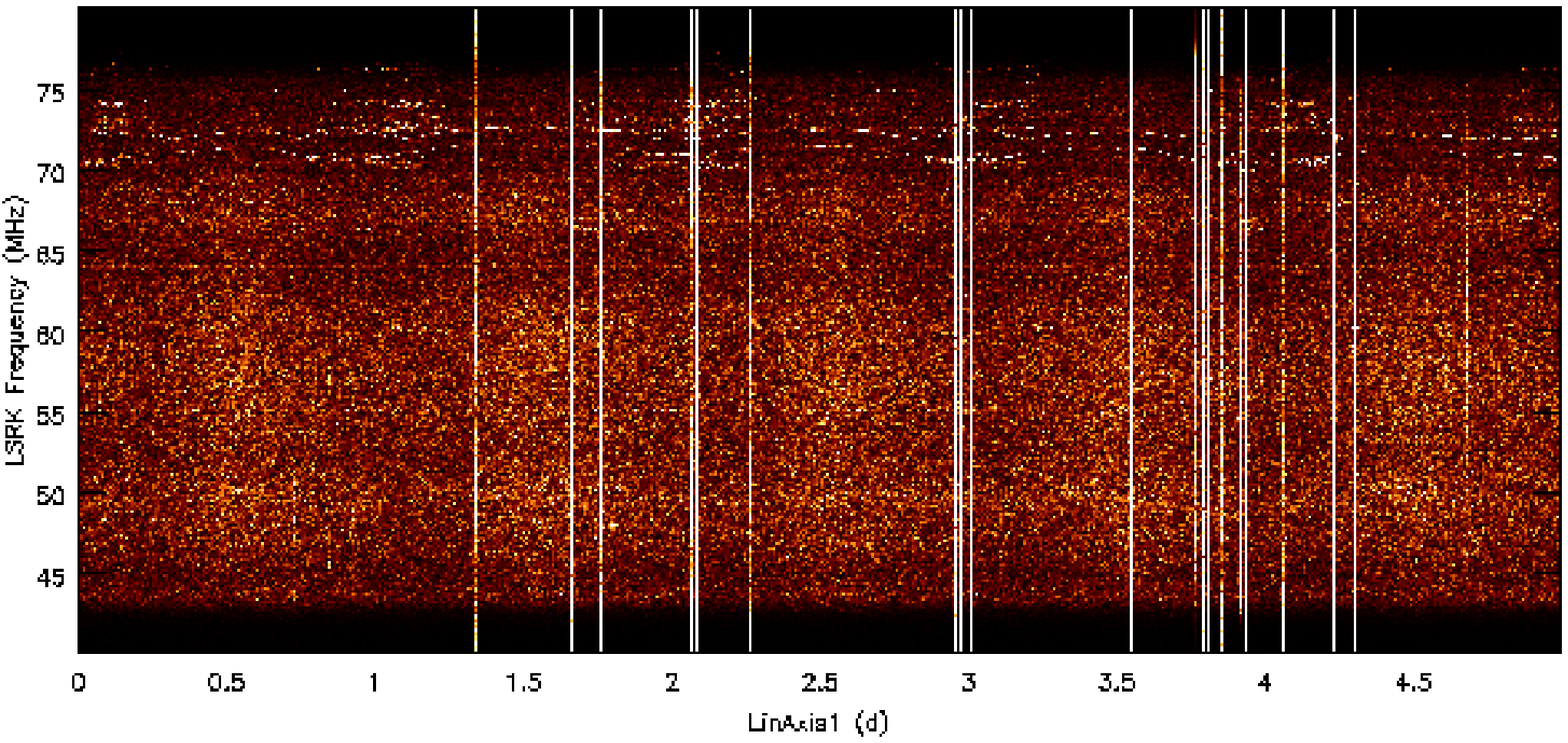}
\includegraphics[width=0.29\linewidth]{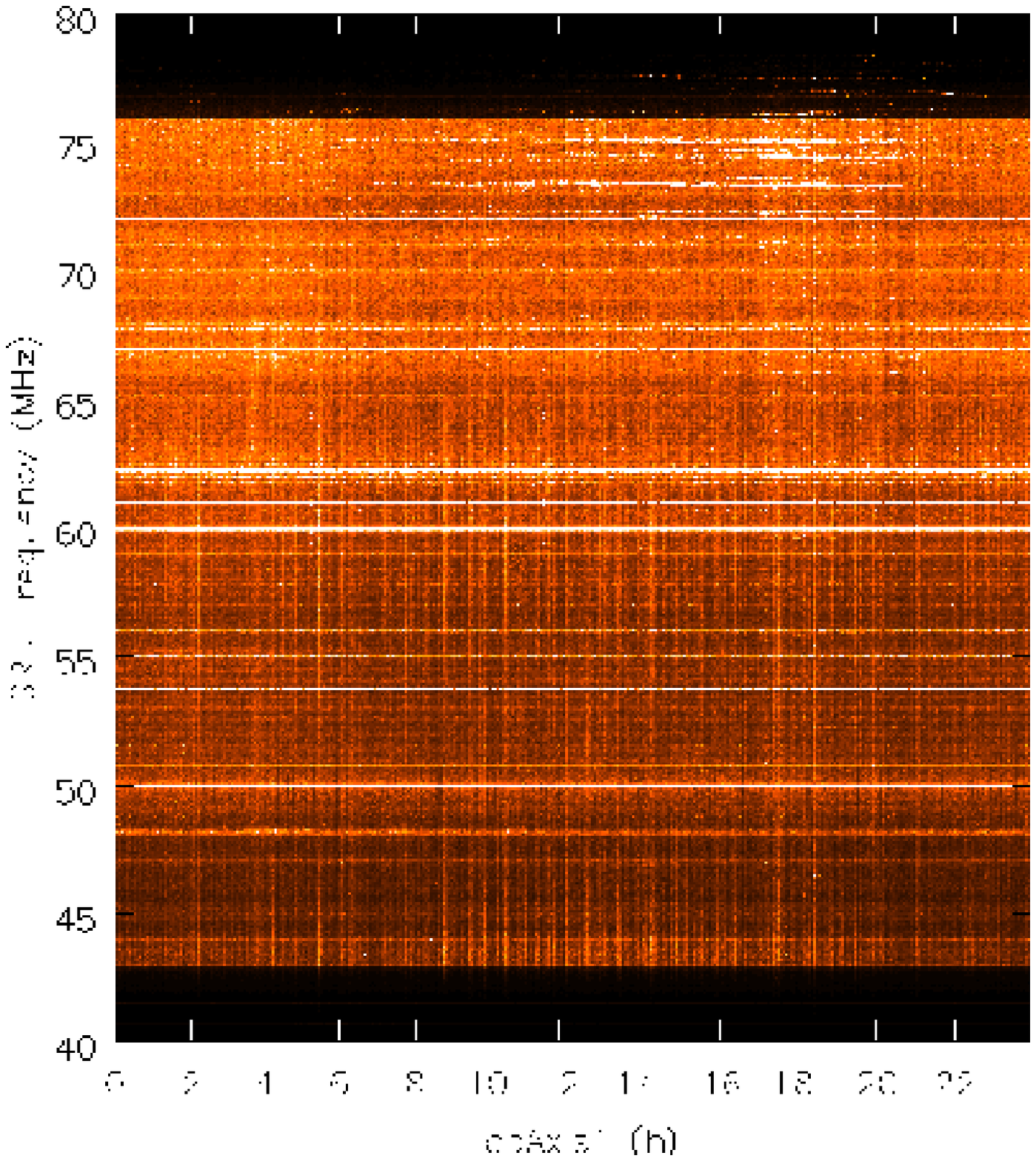}
\caption{
 \label{fig:ft-dwingeloo}
Dynamic spectrum from test measurements in Dwingeloo (left) and KASCADE-Grande 
(right). The y-axis shows the frequency in MHz, the x-axis the time in days 
(left) and hours (right).
The day to night change in general background noise in the Dwingeloo 
measurements is due to the rise and fall of the galactic plane.
Horizontal lines show narrow band interference. Vertical lines are either
due to digital errors or short time pulses.
   }
\end{center}
\end{figure}

During system tests in Dwingeloo some data was taken every minute over a 
period of a few days. From this data we calculated the dynamic spectrum
as shown in figure~\ref{fig:ft-dwingeloo} at the left side. 
The day to night change in general 
background noise is due to the rise and fall of the galactic plane into and 
out of the antenna beam. This shows us, that the noise performance of the total
system is satisfactory.

Horizontal lines in the figure show narrow band interference. E.g. a 
TV-transmitter, that is switched of at night at 62~MHz and 67~MHz.
Vertical lines, that extend over the whole range, are caused by digital errors.
Those, that show the shape of the analog filter are caused by bursts of short
time interference pulses. The vertical line at 3.9~days is caused by a solar 
burst.

Similar measurements at the KASCADE-Grande site (figure~\ref{fig:ft-dwingeloo} 
right) show that there is more radio interference present. This interference 
also prevents us from seeing the rise and fall of the galactic plane. 

\subsection{Astronomic Maps}

\begin{figure}
\begin{center}
\includegraphics[width=0.29\linewidth]{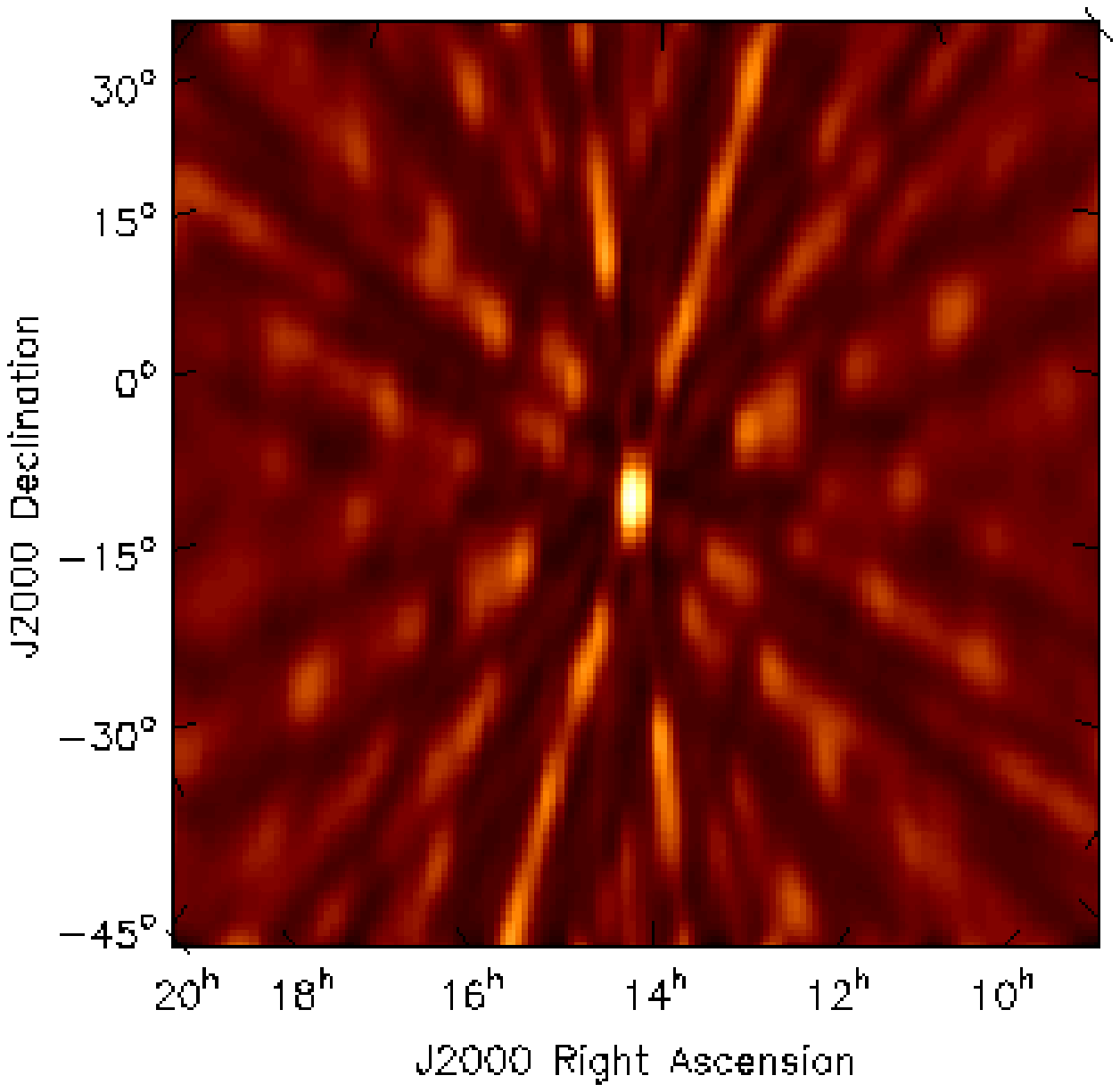}
\includegraphics[width=0.29\linewidth]{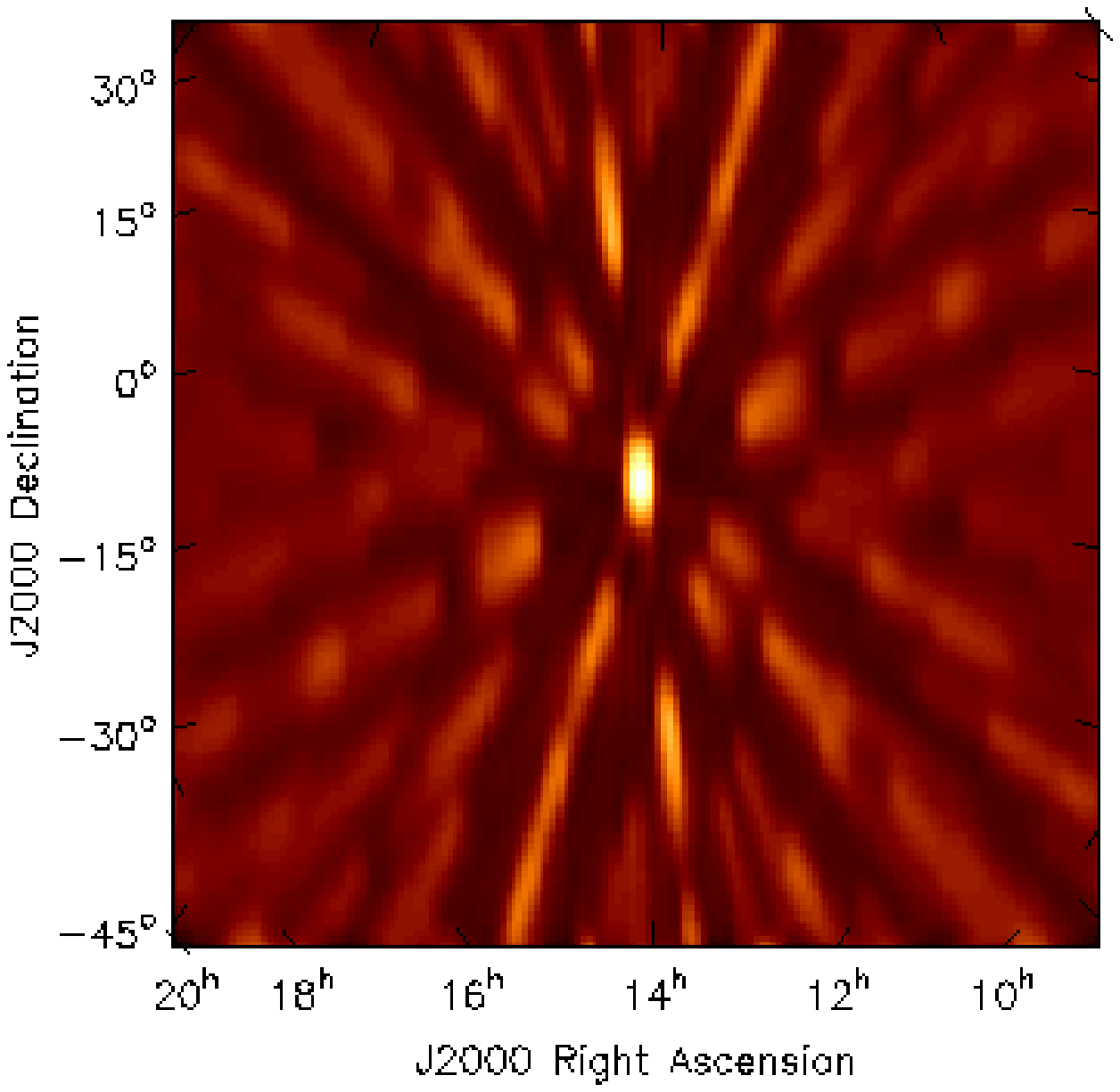}
\includegraphics[width=0.29\linewidth]{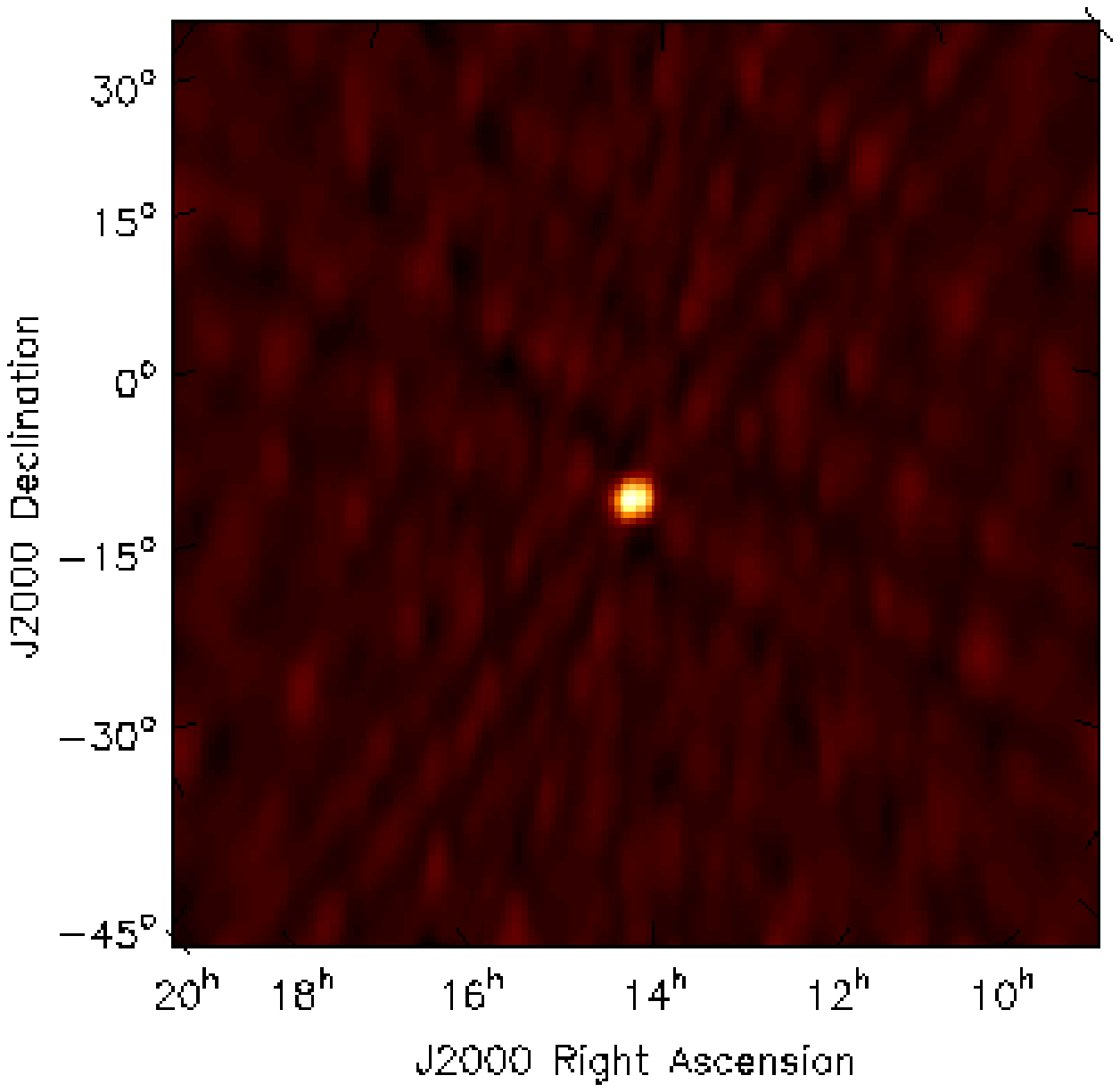}
\caption{
 \label{fig:sun-clean}
Map of the sun during a solar burst. Dirty map (left) in J2000 coordinates and 
SIN projection, simulation of a point source (middle), clean map (right).
   }
\end{center}
\end{figure}

During solar bursts the sun becomes by far the brightest source in the sky at
our frequencies, enabling us to generate an image of the sun with minimal
integration time.
Due to the low frequencies of LOPES the sun is seen as a single
point. This allows us to calibrate the relative time delays of our antennas, by
comparing the measured time delays to the values as expected from the position
of the sun. 

The success of this calibration can be tested by generating a map of the sun.
Figure~\ref{fig:sun-clean} shows a raw (dirty) map from 0.82~milliseconds of
data for a solar burst at the 28 October 2003 on the left side. The structures
around the point in the middle are due to the sidelobes caused by the large 
distances between our antennas. These sidelobes can be computed from the 
antenna positions with standard astronomical software e.g. 
aips++\cite{aips++}
(figure~\ref{fig:sun-clean} middle). With this information one can reconstruct
a clean map of the sun (figure~\ref{fig:sun-clean} right).

By monitoring the relative phases of a TV transmitter we can monitor the phase
stability of our system and get time delay calibration values for every day.
These measurements show us, that sometimes the synchronization of one antenna 
jumps by two sample times\footnote{1~sample time = 1/(80~MHz) = 12.5~ns} for 
some time. But these jumps can be corrected for 
by the data from the TV transmitter. After the correction of integer sized 
jumps the residual time delays are below 0.1~sample time.

\begin{figure}
\begin{center}
\includegraphics[height=0.48\linewidth,angle=-90]{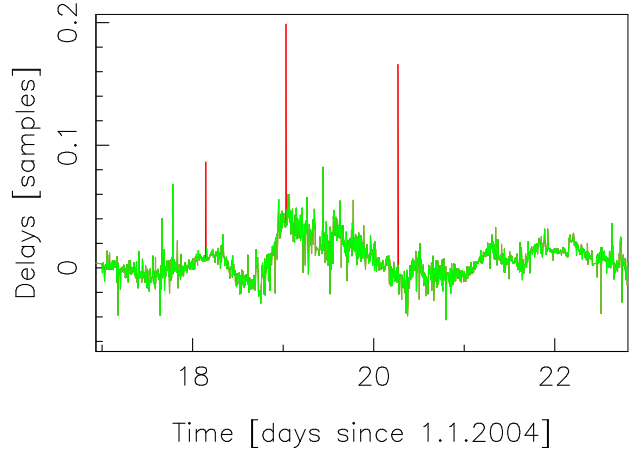}
\caption{
 \label{fig:tvphases}
Time delay deviations for one LOPES antenna, as computed from the relative 
phases of a TV transmitter. The three flagged spikes are where the algorithm 
failed e.g. due to excessive noise.
   }
\end{center}
\end{figure}

\subsection{Air Shower Events}

\begin{figure}
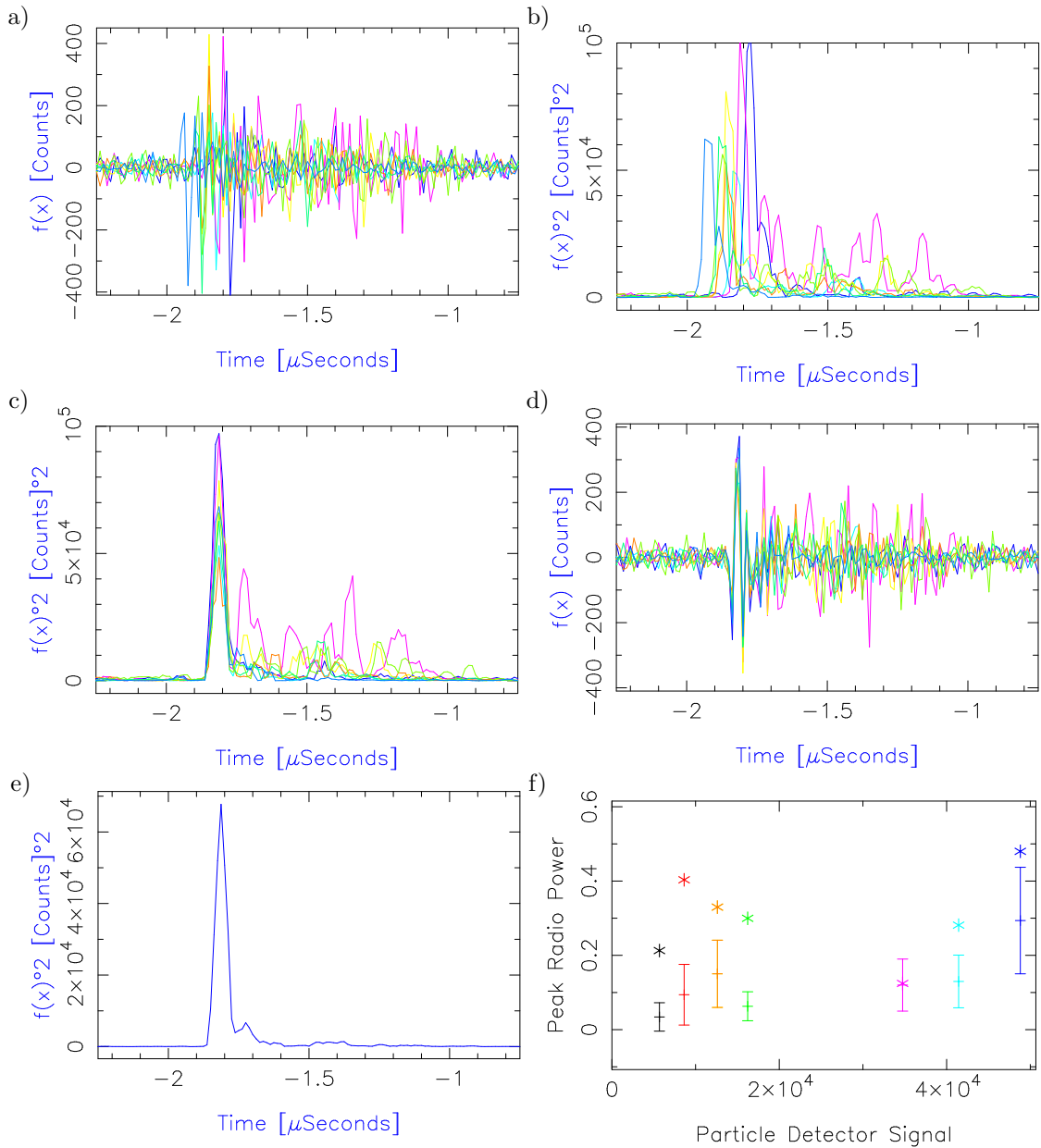

\begin{center}
a)\includegraphics[height=0.41\linewidth,angle=-90]{large-event-e-field-zenith.eps}
b)\includegraphics[height=0.41\linewidth,angle=-90]{large-event-power-zenith.eps}\\
c)\includegraphics[height=0.41\linewidth,angle=-90]{large-event-power.eps} 
d)\includegraphics[height=0.41\linewidth,angle=-90]{large-event-e-field.eps}\\
e)\includegraphics[height=0.41\linewidth,angle=-90]{large-event-power-beam.eps}
f)\includegraphics[height=0.41\linewidth,angle=-90]{EDep-radiomax-TheEvent-singleant-a.ps}
\caption{
 \label{fig:theevent}
Steps of the digital beamforming for a strong air shower event. a) The 
electric field of 8 antennas after filtering of narrow band interference.
b) Received power of those antennas, i.e. the square of the electric field.
c) Received power after time shifting, that is needed to form a beam into the
direction of the air shower. 
d) The electric field after time shifting. 
e) Received power of the formed beam (square of the sum of the values in d). 
The short pulse at ca.~$-1.8~\mu s$ adds up coherently, while the later pulses
(the noise from the particle detectors) do not add up so well.
f) The maximum of the radio peak for the different antennas plotted against the
amount of energy deposited in the detectors around the respective antenna 
(as measured by KASCADE).
Measured value for this air shower event (stars)
and the expected values for interference pulses 
as calculated from the KASCADE data
(error bars).
In all but one antenna the value for this event lies well above what is 
usually produced by the particle detectors.
   }
\end{center}
\end{figure}

As already shown radio interference comes in two kinds:
(approximately) continuous narrow band transmissions and short time pulses.
The narrow band interference can be filtered out without problems by 
transforming the data to frequency space, flagging of the interference and
transforming back. Short time pulses, which are not correlated with an air 
shower, can be identified by their arrival time at the different antennas, and
thus suppressed.

The detectors of the KASCADE-array emit radio pulses, that are correlated 
with the air shower arrival time and direction.
They scale with the particle detector signal, so they are present for the 
largest (and thus interesting) air showers and hard to detect in small air 
showers. The interference pulse each antenna sees is dominated by the 
particle detectors nearest to it.
 
However, the emission from the detectors is
not coherent and can be reduced in the beamforming process. This is demonstrated
in figure~\ref{fig:theevent} a--e. 
During beamforming  a time delay is computed for each antenna from its 
position relative to the phase center and the 
desired direction, e.g. the direction of the air shower as given by 
KASCADE-Grande. 
Then the electric field data (subpanel a) 
is shifted by this delay. Shifting by subpixel values can be done by applying
a phase gradient to the Fourier space data.
A coherent signal that comes from the chosen direction has then in every data
set the same form, like the spike at ca.~$-1.8~\mu s$ in 
subpanel d. Incoherent signals show different forms in the
different antennas, like the pulses at $-1.7~\mu s$--$-1~\mu s$.
(Subpanels b and c in figure~\ref{fig:theevent} show the squared values of the 
data in subpanels a and d. This is a value for the received power.)
Then the shifted electric field data from all antennas is added together. Thus 
coherent signals are enhanced, while incoherent signals are suppressed.
Subpanel e shows the power of the formed beam.
The spike gives a large peak, while the interference from
the particle detectors is reduced, demonstrating that the spike is a radio 
signal, that comes from the chosen direction.

Subpanel f demonstrates that this spike lies well above what is 
usually produced by the particle detectors. By choosing events with the same
amount of energy deposited in the detectors around a given antenna 
(as measured by KASCADE),
one gets a dataset, in which events from small air showers, that had their 
center near the chosen antenna, are accumulated.
In this data the radio signal is dominated by the interference from the
particle detectors.
By averaging over this radio signals, one can get a relation with  which one 
can use the KASCADE data to calculate the expected interference signal.
In subpanel f the data from the example event and the expected data is shown
for seven antennas.

\section{Outlook}

The first phase of LOPES consisting of 10 antennas is running and taking 
scientific data.
Currently we are working on the scientific analysis of the data. The results 
of this work will be published in another paper.
The second stage of LOPES with
20 more antennas is currently under construction. It will be set up this year.

The same technology can be applied to other forthcoming digital radio
telescopes like LOFAR and 
the SKA, providing additional detection area for high
energy cosmic rays.

\acknowledgments     

LOPES is supported by the German Federal Ministry of Education and
Research, under grant No.\,05 CS1ERA/1 (Verbundforschung Astroteilchenphysik).

\bibliography{spie04}   
\bibliographystyle{spiebib}   

\end{document}